\documentclass[12pt]{iopart}
\usepackage{iopams}  
\usepackage{graphicx}

\begin{document}

\title[]{Temperature dependence of the zero-bias anomaly in the Anderson-Hubbard model:  Insights from an ensemble of two-site systems}

\author{R. Wortis}
\author{W.A. Atkinson}

\address{Trent University, 1600 West Bank Drive, Peterborough, ON, K9J 7B8, Canada} 
\ead{rwortis@trentu.ca}
\begin{abstract}
Motivated by experiments on doped transition metal oxides, 
this paper considers the interplay of interactions, disorder, kinetic energy and temperature in a simple system.
An ensemble of two-site Anderson-Hubbard model systems has already been shown to display a zero-bias anomaly\cite{Wortis2010} which shares features with that found in the two-dimensional Anderson-Hubbard model.\cite{Chiesa2008}
Here the temperature dependence of the density of states of this ensemble is examined.
In the atomic limit, there is no zero-bias anomaly at zero temperature, but one develops at small nonzero temperatures.
With hopping, small temperatures augment the zero-temperature kinetic-energy-driven zero-bias anomaly, while at larger temperatures the anomaly is filled in.
\end{abstract}


\maketitle

\section{Introduction}

Many of the systems which motivate the study of strong electronic correlations display their most interesting behaviors when doped.
Doping introduces disorder, and the role played by this disorder is often unclear.
High temperature superconductors are an iconic example.
Moreover, experiments are performed at nonzero temperature, adding another energy scale to the problem.
The recent proposal of a many-body localization transition at nonzero temperature\cite{Basko2006a} underscores the richness of this phase space.
The goal of this paper is to explore the interplay of interactions, disorder, kinetic energy and temperature in a very simple system.

The phenomenon on which we focus is suppression of the density of states (DOS) near the Fermi level.
This can be seen directly in measurements such as scanning tunneling spectroscopy and is also reflected in transport measurements such as the low temperature conductivity.
Disorder-induced zero-bias anomalies (ZBA) have been studied in many doped transition metal oxides including the manganites,\cite{Sarma1998,Mitra2005} the cuprates\cite{Naqib2005} and the ruthenates.\cite{Maiti2007}
It has been suggested that disorder may stabilize the pseudogap in cuprates,\cite{Chiesa2008}
and the temperature dependence of the ZBA in the manganites may carry information about the metal insulator transition.\cite{Mitra2005}
The form of the DOS in disordered interacting systems is best understood in two limits:
Altshuler-Aronov ZBAs in metallic systems with weak interactions and weak disorder,\cite{Altshuler1985} and the Efros-Schklovskii Coulomb gap in the atomic limit.\cite{Efros1975}
However, DOS measurements on strongly disordered transition metal oxides are not well described by either of these pictures.
A fresh angle on these systems was provided by recent numerical results for the Anderson-Hubbard model.\cite{Chiesa2008,Shinaoka2009,Shinaoka2009b}

The Anderson-Hubbard model (AHM) is among the simplest models to combine disorder and strong correlations.
The Hamiltonian is
\begin{equation}
\hat H = \sum_{i,j,\sigma} t_{ij} \hat c^\dagger_{i\sigma} \hat c_{j\sigma}
+ \sum_i \left (
\epsilon_i \hat n_i + U \hat n_{i\uparrow}\hat n_{i\downarrow}
\right ).
\label{eq:H}
\end{equation}
$t_{ij}=-t$ for nearest-neighbor sites $i$ and $j$, and is zero otherwise. 
$\hat c_{i\sigma}$ and $\hat n_{i\sigma}$ are the annihilation and number operators for lattice site $i$ and spin $\sigma$.
$\epsilon_i$, the energy of the orbital at site $i$, is chosen from a uniform distribution $\epsilon_i \in [-\frac{\Delta}{2},\frac{\Delta}{2}]$.

We consider here an ensemble of two-site systems, referred to as molecules.
It has been shown that the density of states (DOS) of this ensemble displays a zero-bias anomaly (ZBA) generated by a mechanism unique to strong correlations.\cite{Wortis2010,Chen2010}
Moreover, at zero temperature the parameter dependence of this ZBA shares many features\cite{Wortis2010} with that found in exact diagonalization and quantum Monte Carlo studies of the two-dimensional lattice\cite{Chiesa2008} in the limit of large disorder.

Here we present the temperature dependence of the ZBA in the ensemble of molecules.
In the atomic limit, there is no ZBA at $T=0$, but nonzero temperature creates one.
With hopping, the $T=0$ ZBA is initially enhanced as temperature is increased, and then filled in, resulting in non-monotonic temperature dependence of the DOS at the chemical potential, $\rho(\mu)$.
We begin by outlining the features of the ensemble DOS at zero temperature, to provide context.
The nonzero temperature results are then presented and discussed.

\section{Ensemble of molecules at zero temperature}

\subsection{Atomic limit}
\begin{figure}
\begin{center}
\includegraphics[width=3 in]{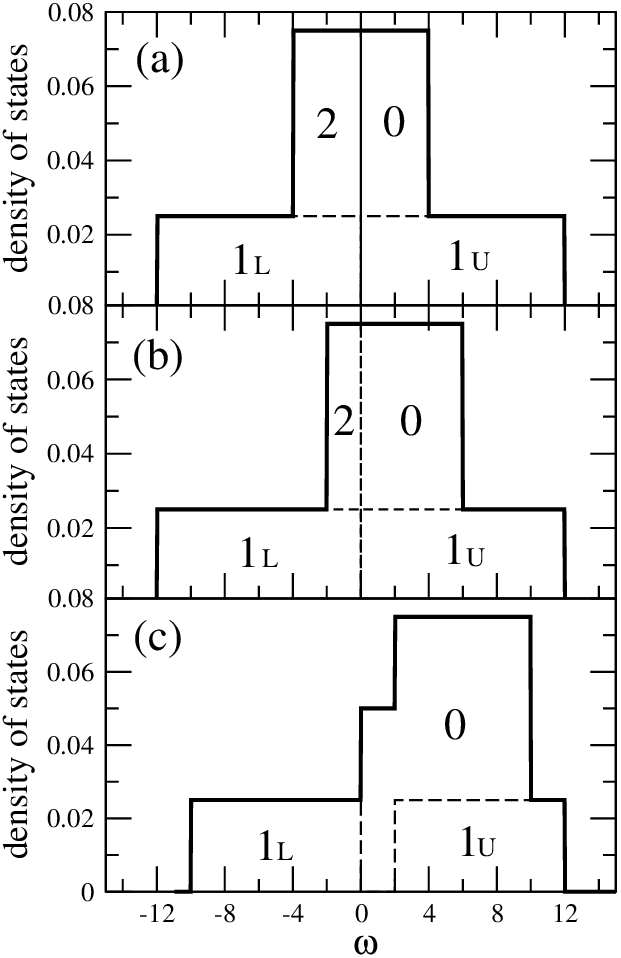}
\caption{(color online) The zero temperature atomic DOS of the AHM for $U=12$, $\Delta=20$ and three dopings:  (a) $\mu=U/2$, (b) $\mu=U/3$ and (c) $\mu=0$.  Labels indicate the ground-state particle number of the sites contributing in each region.  L and U refer to the LHO and UHO contributions respectively.} 
\label{fig:1}
\end{center}
\end{figure}

In general, disorder reduces the importance of kinetic energy relative to interactions, and in the limit of strong disorder 
hopping causes only minimal changes to the atomic spectrum.
In the atomic limit of the AHM, there is no distinction between an ensemble of molecules and a lattice of any dimension.
In this limit, each site contributes to the DOS at, at most, two energies.
If a site with potential $\epsilon$ is empty, it will contribute at energy $\epsilon$, corresponding to adding a particle to the lower Hubbard orbital (LHO).
If it is doubly occupied, it will contribute at energy $\epsilon+U$, corresponding to removing a particle from the upper Hubbard orbital (UHO).
And if it is singly occupied, it will contribute with half the weight at both of these energies.
The resulting atomic DOS for an ensemble of molecules is sketched in Figure \ref{fig:1} at three different fillings.
Panel (a) is at half filling and the DOS is symmetric with a high plateau in the centre and lower shoulders.  In panel (b), the system is underdoped, but the Fermi level remains in the central plateau.  In panel (c), the system is further underdoped and the Fermi level sits at a step edge.  The transition from configuration (b) to (c) occurs at $\mu-U=-\Delta/2$, and the Fermi level remains at a step edge down to $\mu=-\Delta/2$, below which $\rho(\mu)=0$.

\subsection{Nonzero hopping}
At zero temperature, the poles in the molecular Green's function depend on the ground state of the molecule.
For all ground states, the addition of hopping shifts the positions of these poles.
The largest contribution to the ZBA comes from molecules for which (i) there are two particles in the ground state and (ii) the LHO on one site is close in energy to the UHO on the neighboring site.\cite{Wortis2010}
The origin of the anomaly can be seen as follows:
While one of the two electrons in the ground state sits almost entirely on the lower energy site, the second electron is spread between the UHO on this site and the LHO of the neighboring site.
This spreading reduces the kinetic energy of the ground state by an amount of order $t$.
Transitions from the ground state are correspondingly increased in energy.
That this causes a reduction in the DOS specifically near the Fermi level can be seen from the fact that this particular arrangement with two particles in the ground state can only occur when the alignment of LHO and UHO occurs near $\epsilon_F$.
Note that this mixing specifically of LHO and UHO on neighboring sites is unique to strongly correlated systems $U/t \gg 1$.
Additional contributions coming from other molecular ground states have been explored.\cite{Chen2010}

We are motivated by a desire to understand the ZBA found in two-dimensional lattices.\cite{Chiesa2008}
Given that there is no ZBA in the atomic limit, a molecule is the smallest AHM system to display ZBA physics at zero temperature.
In considering an ensemble of molecules, we can expect to capture aspects of the physics in strongly disordered lattices.
Because the ensemble of molecules has only short length scales, its behavior at low energies will differ from that of the lattice.
Focusing, therefore, on the width of the ZBA, we found when $\Delta>U$ the width is of order $t$ and independent of $U$, $\Delta$ and doping.
This is consistent with results on two-dimensional lattices for disorder much larger than the band width and $U \gg t$.

\section{Results at nonzero temperature}

\subsection{Atomic limit}

\begin{figure}
\begin{center}
\includegraphics[width=5 in]{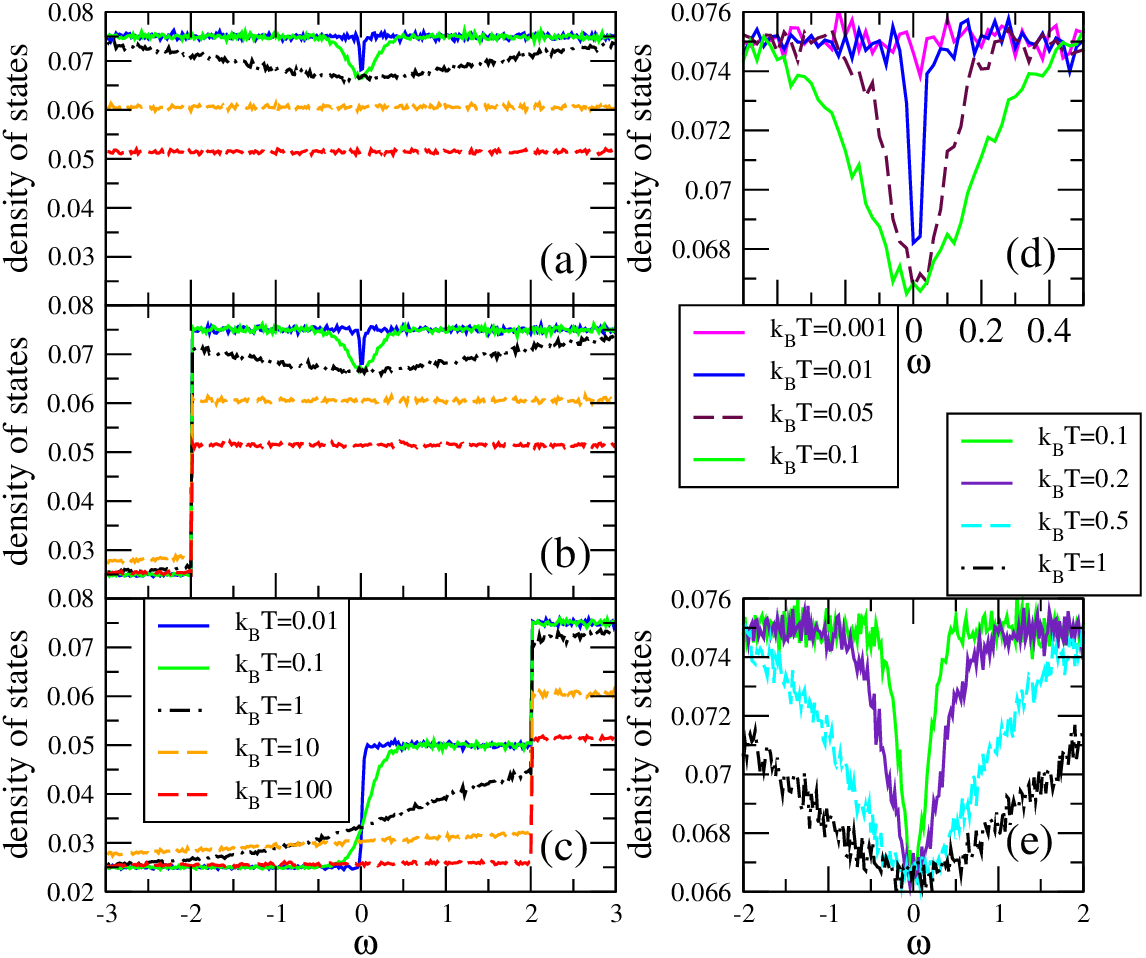}
\caption{(color online) Temperature dependence of the atomic AHM DOS near the Fermi surface for $U=12$, $\Delta=20$ and three dopings:  (a) $\mu=U/2$, (b) $\mu=U/3$ and (c) $\mu=0$.  Panels (d) and (e) focus on the temperature induced ZBA at half filling.  The ensemble includes 10 million molecules.} 
\label{fig:2}
\end{center}
\end{figure}

Figure \ref{fig:2} shows the variation of the atomic DOS at three different doping levels corresponding to those in Figure \ref{fig:1}.
In panels (a) and (b), small nonzero temperatures generate a ZBA.  Examination of panels (d) and (e) show that the width of the anomaly grows roughly linearly with $k_B T$.

The origin of this anomaly is as follows.  Consider a site with a potential just above the chemical potential, $\epsilon_0 \gtrsim \mu$.  
At zero temperature it is empty, and contributes to the DOS at $\epsilon_0$.
When the temperature is not zero, the site will be singly occupied with Boltzmann probability $\propto e^{-(\epsilon_0-\mu)/k_BT}$.  
This causes some of the spectral weight for this site to shift from $\epsilon_0$ to $\epsilon_0+U$.  
Similarly, for sites with energies $\epsilon_2$ just below $\mu-U$, which are doubly occupied at zero temperature, nonzero temperature causes spectral weight to shift from $\epsilon_2+U$ to $\epsilon_2$, again moving spectral weight away from the chemical potential.
The effect of sites which are singly occupied at zero temperature is the reverse, increasing spectral weight at $\mu$.  
In panels (a) and (b) the contribution to the DOS at the chemical potential from empty and doubly occupied sites is greater than that from singly occupied sites.
Therefore, a negative anomaly forms.  
In contrast, in panel (c) nonzero temperature increases the DOS below $\epsilon_F$ and reduces it above, smoothing the step edge.
Note that although the width of this anomaly is proportional to $T$, the energy scale over which spectral weight is transferred is $U$.
The states removed from the Fermi level do not pile up at the edge of the anomaly, as they do, for example, in the kinetic-energy-drive ZBA.\cite{Wortis2010}

As the temperature is increased further, all sites begin to have contributions at both $\epsilon$ and $\epsilon + U$.
In the high temperature limit, the maximum value of the DOS is reduced from ${3 \over 2\Delta}$ to ${1 \over \Delta}$, which is the noninteracting value.

\subsection{Nonzero hopping}

\begin{figure}
\begin{center}
\includegraphics[width=5 in]{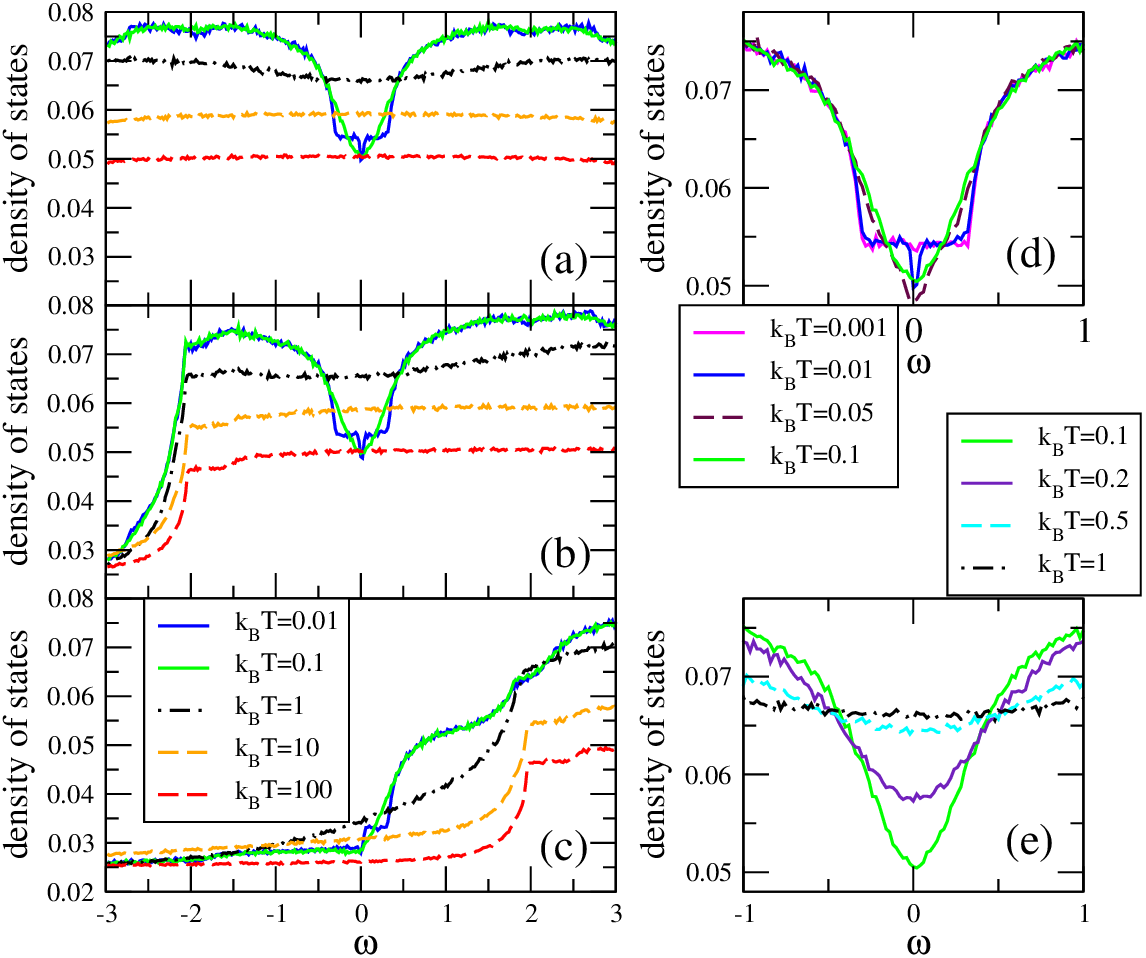}
\caption{(color online) The temperature dependence of the DOS near the Fermi surface of an ensemble of two-site AHMs.  Parameters are as in Figure \ref{fig:2} with $t=1$.} 
\label{fig:3}
\end{center}
\end{figure}

Figure \ref{fig:3} shows the variation with temperature of the DOS at the same three fillings when $t\ne0$.
In the high temperature limit, the DOS with hopping is very similar to that in the atomic limit.
Each molecule has four poles of equal weight.  Hopping shifts these poles relative to their atomic positions, rounding the step edges in the atomic DOS, but only slightly so long as $\Delta \gg t$.

At zero temperature, however, hopping creates a ZBA.
Raising the temperature initially adds the temperature-driven anomaly to this kinetic-energy-driven one.  
The magnitude of the temperature-driven addition to the ZBA is only slightly reduced from its atomic limit magnitude.
This is because there is only partial overlap between the molecules responsible for the temperature-driven anomaly and the kinetic-energy-driven one.
The kinetic-energy-driven ZBA comes from molecules in which one site potential is near $\mu$ and the other is near $\mu-U$.
The temperature-driven ZBA comes from molecules for which at least one site potential is near $\mu$.  
This includes all the molecules responsible for the kinetic-energy-driven ZBA but also many more.

As temperature is further increased, the molecules responsible for the kinetic-energy-driven ZBA have significant probability of {\em not} occupying the special ground state which is lower in energy by $t$.
As a result, this anomaly fills in from the bottom up.
The result is a very non-monotonic temperature dependence of $\rho(\mu)$:  It first declines as temperature is increased from zero, due to the addition of the temperature-driven ZBA.  
It then rises as the kinetic-energy-driven ZBA is washed out.
And finally it falls again as the high temperature limit is approached.


These results may shed light on existing experiments.
Figure \ref{fig:3} (e) resembles the temperature dependence shown in Figure 2 of Reference \cite{Sarma1998}, showing the ZBA in LaNi$_{0.8}$Mn$_{0.2}$O$_3$ filling in when the temperature is increased.  Moreover, measurements on La$_{0.7}$Ca$_{0.3}$MnO$_3$ have observed a ZBA the depth of which has a similar non-monotonic dependence on temperature.\cite{Mitra2005}

In conclusion, we have explored the temperature dependence of the ZBA in the AHM in the context of an ensemble of two-site systems.
We find a temperature-driven ZBA in the atomic limit.
When hopping is nonzero, a temperature-driven ZBA enhances the kinetic-energy-driven ZBA at low temperatures, while at higher temperatures the kinetic-energy-driven ZBA fills in from the bottom.
This enhancement of the ZBA at low temperatures is in marked contrast to weakly interacting systems in which the Altshuler-Aronov anomaly is weakened at any nonzero temperature.

\section*{References}

\end{document}